\documentclass[aps,prx,showpacs,twocolumn,amsmath,amssymb,superscriptaddress,letterpaper]{revtex4}
\usepackage{times}
\usepackage{amsfonts}
\usepackage{mathrsfs,ulem}
\usepackage{graphicx}
\usepackage{dcolumn}
\usepackage{bm}
\usepackage{color,xcolor}

\usepackage[colorlinks,bookmarks=false,citecolor=blue,linkcolor=red,urlcolor=blue]{hyperref}
\usepackage{hyperref}

\newcommand{\ket}[1]{| #1 \rangle}

\newcommand{\be}{\begin{equation}}
\newcommand{\ee}{\end{equation}}
\newcommand{\bma}{\begin{pmatrix}}
\newcommand{\ema}{\end{pmatrix}}
\newcommand{\balig}{\begin{align}}
\newcommand{\ealig}{\end{align}}

\newcommand{\bZ}{\mathbb{Z}}
\newcommand{\bI}{\mathbb{I}}
\newcommand{\ba}{\begin{eqnarray}}
\newcommand{\ea}{\end{eqnarray}}

\newcommand{\ignore}[1]{}

\newcommand{\bk}{{\bm{k}}}

\newcommand{\br}{\bm{r}}
\newcommand{\bu}{\bm{u}}
\newcommand{\bv}{\bm{v}}

\begin{document}

\title{Yu-Shiba-Rusinov states in a superconductor with topological $\bZ_2$ bands}

\author{Ching-Kai Chiu}
\affiliation{RIKEN Interdisciplinary Theoretical and Mathematical Sciences (iTHEMS), Wako, Saitama 351-0198, Japan}
\affiliation{Kavli Institute for Theoretical Sciences, University of Chinese Academy of Sciences, Beijing 100190, China}

\author{Ziqiang Wang}
\affiliation{Department of Physics, Boston College, Chestnut Hill, Massachusetts 02467, USA}

\date{\today}

\begin{abstract}
A Yu-Shiba-Rusinov (YSR) state is a localized in-gap state induced by a magnetic impurity in a superconductor. Recent experiments used an STM tip to manipulate the exchange coupling between an Fe adatom and the ${\rm FeTe}_{0.55}{\rm Se}_{0.45}$ superconductor possessing a $\bZ_2$ nontrivial band structure with topological surface states.  As the tip moves close to the single Fe adatom, the energy of the in-gap state is modulated and exhibits two unusual zero-energy crossings, which cannot be understood by coupling the magnetic impurity to the superconducting topological surface Dirac cone.
Here, we numerically and analytically study the YSR states in superconductors with nontrivial $\bZ_2$ bands and explain the origin of the two zero-energy crossings as a function of the exchange coupling between the magnetic impurity and the {\it bulk} states.  We analyze the role of the topological surface states and compare in-gap states to systems with trivial $\bZ_2$ bands. The spin-polarization of the YSR states is further studied for future experimental measurement.
\end{abstract}

\maketitle

A magnetic impurity adatom deposited on a superconductor can induce an in-gap bound state, which is known as a Yu-Shiba-Rusinov (YSR) state~\cite{Yu_magnetic,Shiba:1968vm,Rusinov_magnetic}. It has been theoretically proposed that the arrangement of the YSR states on the superconductor surface can lead to topological superconductor platforms hosting Majorana zero modes~\cite{PhysRevX.9.011033,PhysRevB.93.140503,PhysRevLett.111.206802,
PhysRevB.90.085124,PhysRevB.88.180503,Chiu_2018,Li:2016wx,PhysRevB.88.155420,PhysRevLett.111.186805,Pawlak:2016tb,WrayCavaHasan,PhysRevLett.107.217001}. The experimental progress serves as hints of the Majorana existence~\cite{Fan:2021td,Kimeaar5251,Palacio-Moraleseaav6600,Nadj-Perge_Ferro_SC}. Understanding the fundamentals of the YSR state~\cite{PhysRevB.92.064503,Kim:2020ut,Beck:2021uj} is an important step toward understanding local excitations in superconductors and topological superconductivity. 

Novel properties of the YSR states have been probed by a controlled approach of the STM tip to the magnetic FeP molecules adsorbed on the surface of conventional Pb($111$) superconductors ~\cite{PhysRevLett.121.196803}.
Modulating the tip distance by changing the conductance $G$ of the tunnel barrier (Fig.~1(a)) can directly manipulate the exchange coupling of the magnetic FeP to the superconductor.
Two consecutive zero-energy crossings were observed with increasing $G$~\cite{PhysRevLett.121.196803}. Since the exchange of quasiparticle and quasihole occurs at zero energy, a zero-energy crossing corresponds to a first-order quantum phase transition. However, since the distance between the STM tip and FeP turned out to not be non-monotonic~\cite{PhysRevLett.121.196803},
the two crossings may correspond to the same exchange coupling at a single transition.
Recently, magnetic Fe adatoms deposited on the surface of  Fe-based superconductor ${\rm FeTe}_{0.55}{\rm Se}_{0.45}$ were studied by STM using a similar technique~\cite{Fan:2021td}. A zero-energy crossing of the YSR states with increasing $G$ was observed and accompanied by a monotonic decrease in the tip distance $d$ to the magnetic Fe adatom. Surprisingly, further increasing the exchange coupling by reducing $d$ causes the YSR states to reverse trajectories and return to zero energy,
in striking contrast to the expected behavior in conventional superconductors~\cite{Yu_magnetic,Shiba:1968vm,Rusinov_magnetic}.
The emergence of the zero-energy bound state robust against further reduction of $d$ was interpreted as evidence for the transition to the quantum anomalous vortex spontaneously nucleated at magnetic Fe impurities~\cite{PhysRevX.9.011033},  trapping a vortex Majorana zero mode from the SC topological surface states~\cite{Fan:2021td}.
What drives the return of the YSR states to zero energy in a tendency to form a possible second zero-energy crossing was not understood.

In this work,
we study the fundamental reason for the emergence of the two zero-energy crossings of the in-gap YSR state as a function of the exchange coupling in connection to ${\rm FeTe}_{0.55}{\rm Se}_{0.45}$.
The non-trivial topological physics of ${\rm FeTe}_{0.55}{\rm Se}_{0.45}$ stems from the surface Dirac cone due to the topological $\bZ_2$ bands, which becomes superconducting (SC) below the bulk $T_c$ and can host Majorana zero modes in magnetic vortices~\cite{Fu:2008fk,Zhang2018Observation,Wang2018Evidence,
2018arXiv181208995M,Chiueaay0443,Kong:2021wr}. In the effective surface-only theory, where the deposited magnetic impurity exchange couples to the SC Dirac cone exclusively, the in-gap YSR state exhibits only one zero-energy crossing~\cite{PhysRevB.90.125443}, similar to the case of a conventional superconductors~\cite{Yu_magnetic,Shiba:1968vm,Rusinov_magnetic}. The bulk physics and the possible emergence of another YSR state from the topological $\bZ_2$ bands has not been studied.

We focus on the basic physics of the $\bZ_2$ nontrivial band structure, by assuming the effects of the other bulk bands are small or at high energies. We thus consider a microscopic theory where the impurity spin is exchange coupled to the 3D $\bZ_2$ bulk bands, and study the resulting localized in-gap states at the impurity. This procedure emphasizes that the interaction of the impurity spin with the SC topological surface states originates from its exchange coupling to the topological nontrivial bulk bands. Moreover, it captures the possible bulk physics that would be inaccessible in a surface-only effective model.
We numerically and analytically study the entire 3D system modeled after  ${\rm FeTe}_{0.55}{\rm Se}_{0.45}$ and reveal that the second zero-energy crossing comes from the bulk physics.
We study and propose spin-polarized STM as a probe to distinguish the two zero-energy crossings. 



\begin{figure}[!tbp]
  \centering
\includegraphics[width=0.45\textwidth]{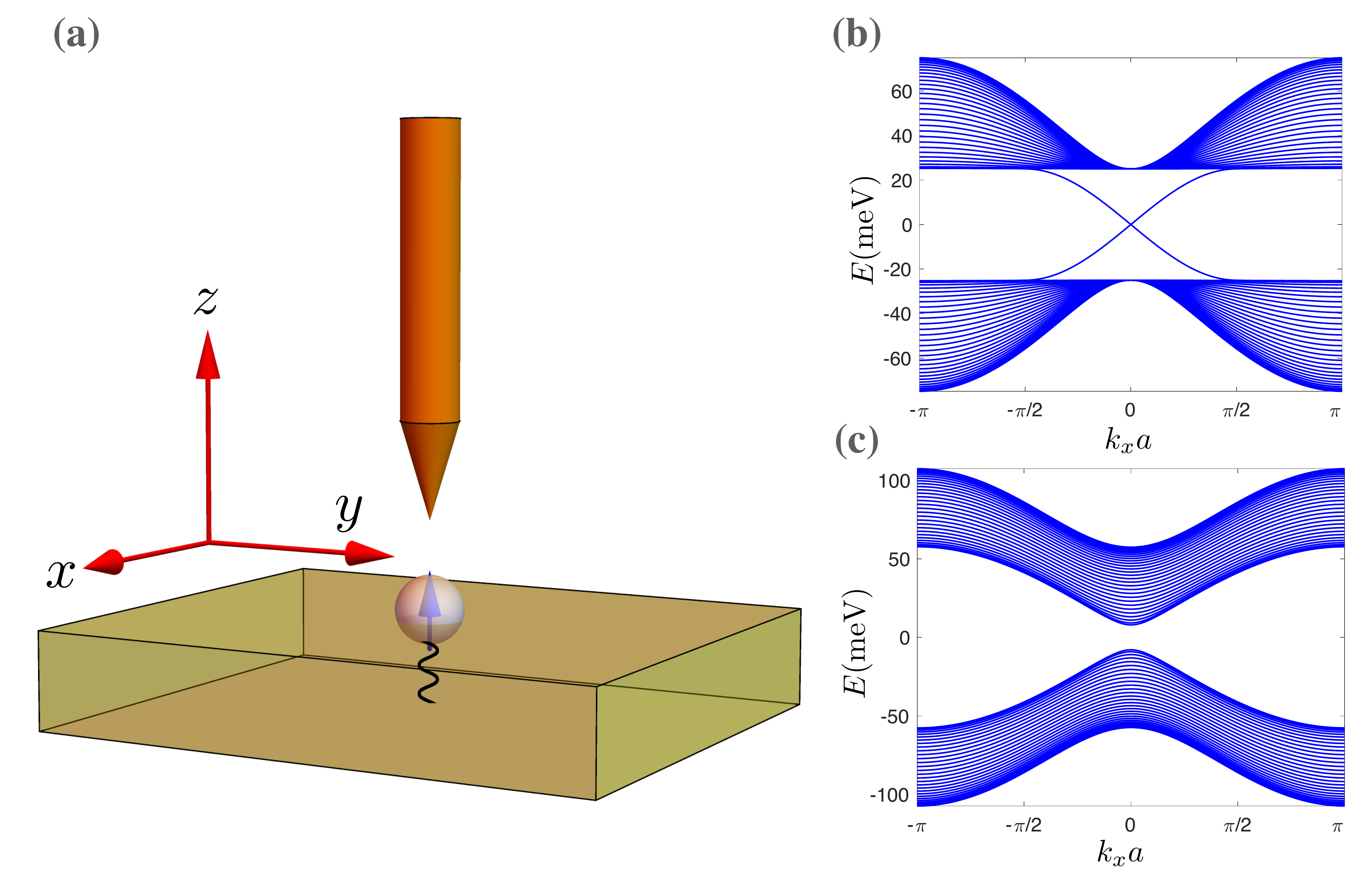}
  \caption{ (a) Schematic setup indicating an Fe adatom on the top surface of ${\rm FeTe}_{x}{\rm Se}_{1-x}$. The STM tip can be moved to approach the adatom, probing the tunneling current and manipulating the exchange coupling. (b) Bulk-surface spectrum at $k_y=0$ for $H_{\rm{TI}}$, showing  the presence of the surface Dirac cone inside the $\bZ_2$ nontrivial TI gap on the (001) surface at $m=2$. (c) Spectrum of $H_{\rm{TI}}$ of the $\bZ_2$ trivial insulator at $m=3.3$, showing the bulk gap and the absence of the surface states. }\label{schametics}
\end{figure}

We start with the Hamiltonian in momentum space describing a 3D strong topological insulator (TI)~\cite{Zhang:2009aa} 
\begin{align}
H_{\rm{TI}}(\bk)=&\nu_FM(\bk) \rho_z \sigma_0 -\mu \rho_0 \sigma_0+v_F \sin k_x a \rho_x \sigma_x\nonumber \\
&+ v_F \sin k_y a \rho_x \sigma_y +v_F \sin k_z  a\rho_x\sigma_z, \label{TIH}
\end{align}
where $M(\bk)=-m+\cos k_x a+ \cos k_y a -\cos k_z a$ and the Pauli matrices $\sigma_\alpha$ and $\rho_\beta$ act in the spin and orbital space respectively.
Based on the experimental measurements~\cite{Zhang2018Observation,Wang2018Evidence}, we choose the Fermi velocity
$v_F=25$nm$\cdot$meV/$a$, wavevector $k_F=0.2$nm$^{-1}$, and the chemical potential $\mu=v_F/k_F=5$meV. For simplicity, we consider a simple cubic lattice and choose the lattice constant $a=1$nm in all directions (c.f.~\cite{PhysRevB.79.054503}).
Note that at $m=3$, the bulk gap closes at $Z$ point describing a bulk topological phase transition for ${\rm FeTe}_{0.55}{\rm Se}_{0.45}$. We therefore use $m=2$ to describe the $\bZ_2$ nontrivial bands where the surface Dirac cone emerges and
maximally localizes on the surface, as shown in the bulk-surface spectrum in Fig.~\ref{schametics}(b). The neutral point of the Dirac cone is located at $\bar{\Gamma}:(0,0)$ in the $(001)$ surface BZ, which is consistent with ${\rm FeTe}_{x}{\rm Se}_{1-x}$. For comparison, we will also study the $\bZ_2$ trivial bands at $m=3.3$ shown in Fig.~\ref{schametics}(c) where the gapless surface spectrum is absent.
The superconducting topological surface states, as observed in ${\rm FeTe}_{0.55}{\rm Se}_{0.45}$,  can be generated by introducing bulk s-wave superconductivity. The TI Hamiltonian is thus extended to the Bogoliubov-de Gennes (BdG) Hamiltonian
\begin{eqnarray}
H_{\rm BdG}^0&=&
\left(
\begin{array}{cc}
H_{\rm TI} ({\bk}) & {-i\Delta_0}\rho_0\sigma_y \\
{i\Delta_0}\rho_0\sigma_y& -H_{\rm TI}^*({-\bk}) \\
\end{array}
\right),
\label{HBdG}
\end{eqnarray}
where the value of the
SC gap, $\Delta_0= 1.8$meV, has been measured experimentally ~\cite{Zhang2018Observation,Wang2018Evidence,2018arXiv181208995M,Chen:2018aa} and is used through the manuscript.

We position the Fe adatom at the center of the $(001)$ surface
as shown in Fig.~\ref{schametics}(a). An exchange coupling to the $\bZ_2$ bands is induced at the center of the top surface, denoted as $\br_{top}\equiv{\bf 0}$. Since the lattice translation symmetry is broken, it is necessary to rewrite the BdG Hamiltonian in the real space
\begin{align}
\hat{H}_{\rm{BdG}}&= \sum_{\br}
\bma
c^\dagger_{\br} & c_{\br}
\ema
\bma
H_0(m) & -i\Delta_0\rho_0 \sigma_y \\
i\Delta_0 \rho_0 \sigma_y & -H_0^*(m)
\ema
\bma
c_{\br} \\
c^\dagger_{\br}
\ema  \nonumber \\
&+\sum_{\br,\bm{\delta}}
\bma
c^\dagger_{\br} & c_{\br}
\ema
\bma
H_{\rm{nn}}(\bm{\delta}) & 0 \\
0 & -H_{\rm{nn}}^*(\bm{\delta})
\ema
\bma
c_{\br+\bm{\delta}} \\
c^\dagger_{\br+\bm{\delta}}
\ema,\label{hreal}
\end{align}
where the on-site part $H_0(m)=v_F m \rho_z \sigma_0 - \mu  \rho_o \sigma_0 +
M_z\rho_0 \sigma_z\delta(\br-\br_{\rm top})$ contains the spin exchange coupling $M_z$ taken to be along the $z$-direction first. Here we assume each orbital has the same strength exchange coupling and the supplementary material~\cite{supplement_YSR} shows the simulation for different orbital strengths. In the nearest neighbor hopping part,
$\bm{\delta}=\pm a \hat{x},\pm a \hat{y},\pm a \hat{z}$ indicates the three directions of the hopping, and  $H_{\rm{nn}}(\pm a \hat{n})=-v_F \rho_z \sigma_0/2 \pm i v_F \rho_x \sigma_n/2$ for $n=x,y,z$.


\begin{figure}[!tbp]
  \centering
\includegraphics[width=0.45\textwidth]{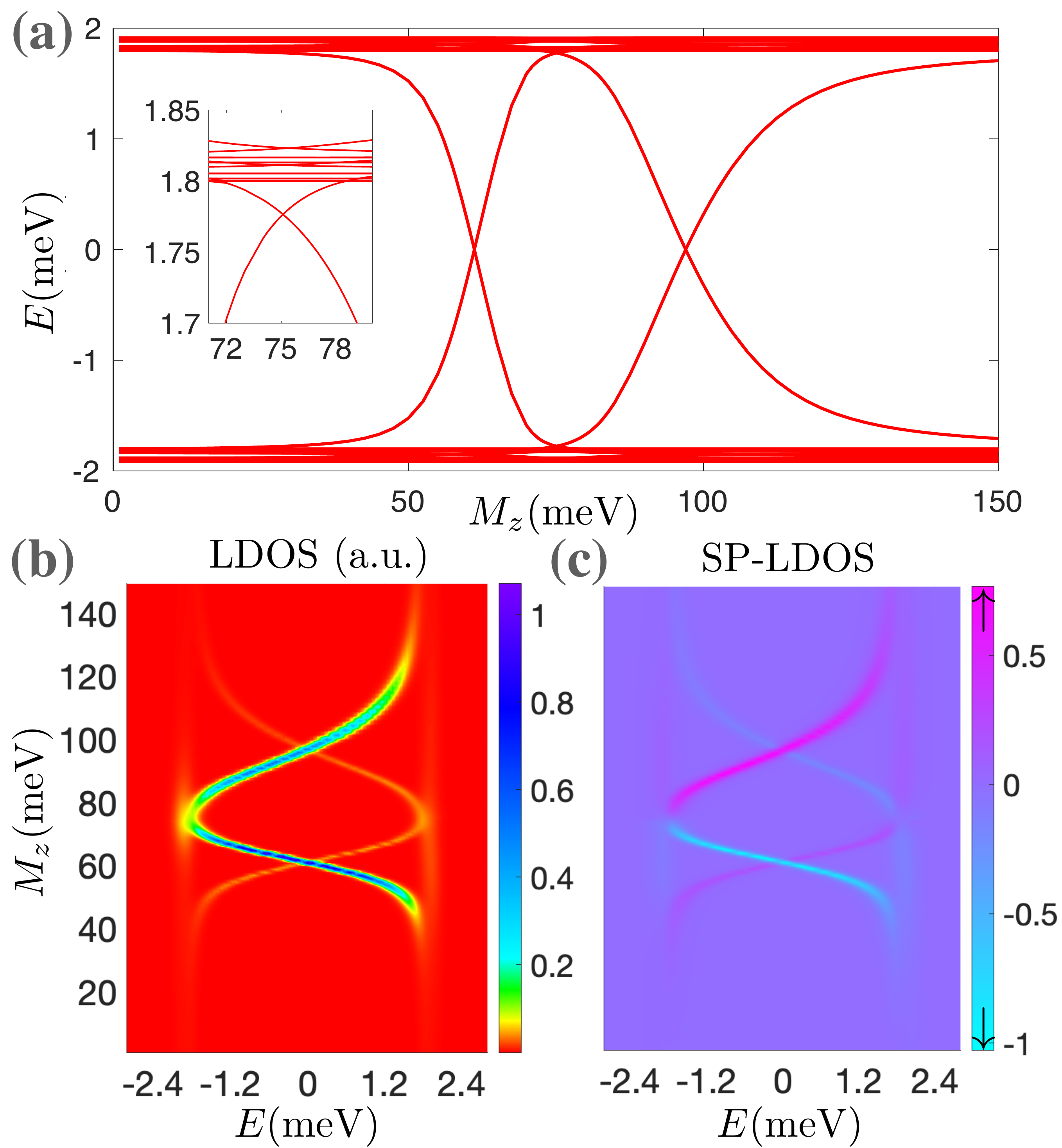}
  \caption{Evolution of the YSR states with increasing exchange coupling ($\mu=5$meV, $m=2$, $M_x=M_y=0$). (a) Spectrum of the in-gap states, showing that YSR states pass through zero-energy twice. The inset shows that the crossing near the gap edge is formed by the first YSR state moving up and a new YSR state moving down in energy. (b) At non-zero exchange couplings, the intensity of the YRS state in the LDOS $N$ breaks particle-hole symmetry. The coherence peaks associated with the SC gap cannot be seen clearly due to the dominating spectral weight of the YSR states. (c) The normalized spin-resolved LDOS $N_p$. The highest peak is spin-down polarized at small to moderate $M_z$, and another main YSR peak with spin-up polarization moves from $-\Delta_{0}$ to $\Delta_{0}$.   }\label{m2}
\end{figure}
	
Since the YSR state is localized on the top surface, the surface size ($L_xL_ya^2$) and the number of layers $L_z$ in the $z$-direction does not significantly affect this in-gap bound state. Hence, we choose $L_x=L_y=80$ and $L_z=5$ for the following simulations. By performing Lanczos algorithm, we find states $\Psi_j(\br)=(\bu_j(\br),\bv_j(\br))$ within and near the SC gap. Fig.~\ref{m2}(a) shows the evolution of the lowest energy eigenstates as the exchange coupling $M_z$ increases from $0$. The trajectories within the superconductor gap ($1.8$meV) corresponds to those of the YSR states localized near the Fe adatom, which indeed exhibit two zero-energy crossings. Fig.~\ref{m2}(b) shows the LDOS at the adatom ($\br={\bf0}$)  given by
\be
N(\br,E)=  \sum_{E_j<0} \Big[ {|\bm{u}_j ({\br}) |^2\over \cosh^2(\beta (E-E_j))}+ {|\bm{v}_j ({\br})  | ^2\over \cosh^2(\beta(E+E_j)) }\Big]
\nonumber
\ee
and $1/\beta=k_BT=0.05$meV.
The spin-resolved LDOS ($N_\uparrow, N_\downarrow$) can be computed by selecting the designated spin part. The normalized spin-polarized (SP) LDOS $N_p=(N_\uparrow-N_\downarrow)/\sqrt{N_\uparrow+N_\downarrow}$  is plotted in Fig.~\ref{m2}(c) at $\br={\bf0}$. Starting from the weak exchange coupling $M_z$, the LDOS of the YSR states exhibits particle-hole symmetric peak energy positions but with asymmetric spectral intensity (Fig.~\ref{m2}b). Fig.~\ref{m2}(c) shows that the highest peak, which is spin-down polarized at positive energy, is much higher than the spin-up polarized second highest peak at negative energy. As $M_z$ increases, these two peaks move close to each other and then cross the Fermi level at zero energy. As the main peak continues to move toward $-\Delta_0$, the second peak moves toward $\Delta_0$.

Surprisingly, contrary to the expected behavior where the two peaks merges into the continuum at $\pm\Delta_0$ with further increasing of the exchange coupling, another YSR state emerges from the continuum and the two distinct YSR states form an energy crossing just inside the SC gap as shown in the inset in Fig.~\ref{m2}(a). Moreover, the newly emerged YSR state carries in Fig.~\ref{m2}(c) an opposite spin polarization  such that the main peak becomes spin-up polarized and moves from $-\Delta_0$ to $\Delta_0$ with increase $M_z$ as can be seen in Fig.~\ref{m2}(b), while the spin-down polarized second peak moves in the opposite direction. This creates a remarkable second zero-energy crossing in the trajectories of the YSR states displayed in Fig.~\ref{m2}.
The presence of the two zero-energy crossings is consistent with the recent experimental observation~\cite{Fan:2021td} and the supplementary materials~~\cite{supplement_YSR} shows the spatial lint cutes at the two crossings. Fig.~\ref{m2} show that the crossings are located at $M_z^{c1}=61$meV and $M_z^{c2}=97$meV, which are close to the estimated exchange coupling from the neutron scattering experiments ($\sim70$meV~\cite{PhysRevLett.108.107002}).

We study the total angular quantum numbers of the YSR states, since in the presence of the spin-orbital coupling the angular momentum and the spin with superconductivity~\cite{PhysRevB.93.174505} are not good quantum numbers separately. The zero-energy crossing of the YSR state reflects the change of the total angular quantum number with its projection on the $z-$axis ($J_z=L_z+S_z$) in the entire system, similar to the first-order transition from a spin-singlet to triplet impurity state in conventional s-wave superconductors~\cite{Yu_magnetic,Shiba:1968vm,Rusinov_magnetic}.  The supplementary material~\cite{supplement_YSR} shows that $J_z$ changes by $-\hbar/2$ as increasing $M_z$ passes through any of the two zero-energy crossings. Hence, for large $M_z$, $J_z$ changes by $-\hbar$ compared to the absence of the Fe adatom.

Each zero-energy crossing comes with a change of fermion parity. At the beginning, the ground state obeys $a_{\rm{YSR}_1}(M_z)\ket{\rm{G}}=0$, $a_{\rm{YSR}_1}(M_z)$ is the annihilation operator of the first YSR state. After the first crossing, the ground state obeys $a_{\rm{YSR}_1}^\dagger(M_z)\ket{\rm{G}}=0$. Therefore, the parity is switched. Similarly, after the second crossing, the parity is changed back to the original one.




\begin{figure}[!t]
  \centering
\includegraphics[width=0.49\textwidth]{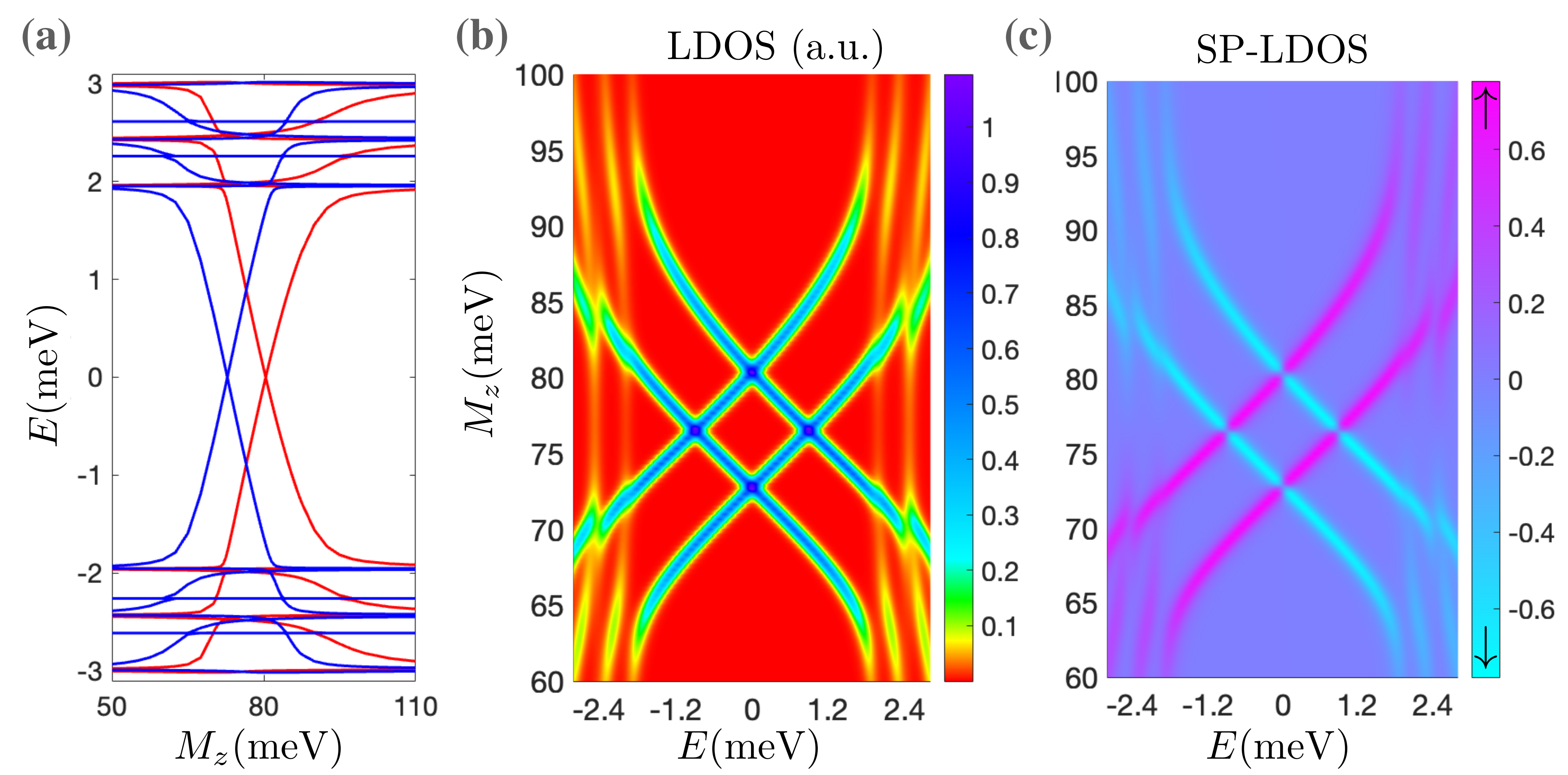}
  \caption{Evolution of the YSR states at zero chemical potential ($\mu=0$, $m=2$, $M_x=M_y=0$). (a) Spectra of the two decoupled Hamiltonian blocks are separately represented by blue lines ($H_+$) and red lines ($H_-$). Each block contributes one zero-energy crossing for $M_z>0$.
 (b) LDOS shows the spectral weight of the YSR states are perfectly particle-hole symmetric. (c) SP-LDOS shares similar features to the case at $\mu=5$meV.  }\label{cp0}
\end{figure}

The presence of two zero-energy crossings hints at the new physics of the YSR states in superconductors with topological $\bZ_2$ bands that is beyond the description by only the SC surface Dirac cone~\cite{PhysRevB.90.125443}. To understand the physical origin of  the two zero-energy crossings, we include the in-plane exchange couplings ($M_x,M_y$) and rewrite the real-space BdG Hamiltonian~(\ref{hreal}) in momentum space by introducing the Pauli matrices $\tau_\alpha$ acting in the particle-hole sector,
\begin{align}
H_{\rm BdG}=& v_F M \tau_z\rho_0 \sigma_0 + v_F \sin k_x \tau_0\rho_x \sigma_x + v_F \sin k_y \tau_z\rho_x\sigma_y  \nonumber \\
& + v_F \sin k_z \tau_0\rho_x \sigma_z + \Delta_0 \tau_y\rho_0 \sigma_y \label{Hmu0} \\
& + \delta(\br) (M_x\tau_z \rho_0 \sigma_x + M_y \tau_0 \rho_0 \sigma_y+ M_z  \tau_z\rho_0\sigma_z), \nonumber 
\end{align}
where the chemical potential $\mu$ has been set to zero for simplicity. Note that $H_{\rm{BdG}}$ has an emergent symmetry and commutes with the operator $P=\tau_y\rho_y\sigma_y$. As a result, we find that the unitary transformation $U=\frac{1}{2\sqrt{2}} (\tau_0-i\tau_x)(\rho_0 - i \rho_x)(\sigma_0 - i \sigma_x)$ simultaneously block-diagonalizes $P$ and $H_{\rm BdG}$. After rearranging the index ($4\rightarrow2,6\rightarrow3,7\rightarrow4,2\rightarrow5,3\rightarrow 6,5\rightarrow7$), we have $UPU^\dagger= {\widetilde P}=\tau_z \rho_0 \sigma_0$, and the block-diagonalized Hamiltonian 
\be
{\widetilde H}_{\rm{BdG}}(\bk)=UH_{\rm BdG}U^\dagger=
\bma
H_+'(\bk) & 0 \\
0 & H_-'(\bk)
\ema	.
\ee
Moreover, we perform another unitary transformation $H_\pm(\bk) = V^\dagger_\pm H_\pm'(\bk)V_\pm$, 
where $V_+=(\tau_z\sigma_0+\tau_x \sigma_0)/\sqrt{2}$ and $V_-=(\tau_0\sigma_0+i\tau_y \sigma_0)/\sqrt{2}$. The explicit forms of the blocks are given by 
\begin{align}
H_\pm(\bk)&= \mp v_F M \rho_y \sigma_y + v_F \sin k_x \rho_0 \sigma_x - v_F \sin k_y \rho_z \sigma_y  \nonumber \\
& \mp v_F \sin k_z \rho_x \sigma_y \pm \Delta_0 \rho_0 \sigma_z\nonumber \\
& + \delta(\br)(\pm M_x \rho_y \sigma_0  \pm M_y \rho_x \sigma_z  - M_z  \rho_z \sigma_z ). \label{Hpm}
\end{align}
First, we only consider the out-of-plane exchange coupling ($M_x=M_y=0$). The numerical result in Fig.~\ref{cp0} shows that the YSR states still exhibit
two zero-energy crossings as a function of $M_z$ in this case. Intriguingly, Fig.~\ref{cp0}(a) reveals that the two crossings respectively stem from the two different Hamiltonian blocks $H_\pm$.
Furthermore, while the spin-polarizations of the YSR states in Fig.~\ref{cp0}(c) are similar to those at $\mu=5$meV displayed in Fig.~\ref{m2}(b), the spectral weights of the LDOS peaks are different and appear perfectly particle-hole symmetric as shown in Fig.~\ref{cp0}(b). This additional symmetry originates from the fact that $P$ connects particle and hole wavefunctions and in the eigen basis of $P$,
$|\bu^\uparrow|^2+|\bu^\downarrow|^2=|\bv^\uparrow|^2+|\bv^\downarrow|^2$.

The block-diagonal Hamiltonian $H_+$ and $H_-$ can be identified as partners of an effective time reversal symmetry $T$. 
This amounts to 
\be
\rho_z \sigma_yH_+^*(-\bk,-M_z)\rho_z \sigma_y=H_-(\bk,M_z) \label{time reversal symmetry}
\ee
such that the YSR state generated by $M_z$ in $H_+$ can be easily transformed to the one generated by $-M_z$ in $H_-$.
 In the supplementary material~\cite{supplement_YSR}, we take the continuum limit of $H_{\pm}$ and provide an approximate, but rather general, analytical solution of the critical exchange couplings for the zero-energy crossings.
For our topological $\bZ_2$ bands at $m=2$, the zero-energy YSR states appear for $H_{\pm}$ at
\be
M_{z,\alpha}^\pm = \pm \frac{ (-1)^{\alpha-1}4\pi v_F^2}{\Delta_0 \ln [1+ \frac{\Lambda^2 v_F^2}{\Delta_0^2+(1-\alpha)v_F^2}]}, \quad \alpha=1,2, \label{m2simple}
%
\ee
where $\Lambda$ is the momentum cutoff.
The order of the exchange couplings corresponding to the zero-energy crossings is given by $M_{z,2}^+<M_{z,1}^-<0<M_{z,1}^+<M_{z,2}^-$. This analytic solution shows that there are two zero-energy crossings for each impurity polarization direction, i.e. each sign of $M_z$, and for $M_z>0$ the first/second crossing belongs to $H_+$/$H_-$, which are consistent with the simulation results shown in Fig.~\ref{cp0}.

Since the YSR states are localized at the Fe adatom on the top surface, additional insights can be gained from the analytical solution. At $m=2$ and on the top surface, the first $2\times 2$ blocks of $H_+$ and $H_-$, which are separately isolated, represent the physics of the SC surface Dirac cone and lead to the zero-energy YSR state at $M_{z,1}^+>0$ and $M_{z,1}^-<0$~\cite{supplement_YSR}. This part corresponds to what is captured in an effective theory with the SC topological surface Dirac cone alone~\cite{PhysRevB.90.125443}. The second zero-energy crossing at $M_{z,2}^->0$ and $M_{z,2}^+<0$ emerges from the remaining block of the Hamiltonian and involves the coupling of the magnetic impurity to the bulk bands at the surface. This is ultimately related to the strongly spin-orbit coupled two-orbitals of the nontrivial $\bZ_2$ band structure with band inversion. We have verified that the two zero-energy crossings of the YSR states as a function of the exchange coupling remain robust when the Fermi level crosses the bulk band~\cite{supplement_YSR}.
For completeness, we have studied the model at $m=3.3$ (Fig.~\ref{schametics}(c)), where the $\bZ_2$ bands are topological trivial with the absence of band inversion and Dirac cone surface states, and found very different behaviors both numerically and analytically~\cite{supplement_YSR}. Unlike YSR states with a particle or hole counterpart, as the chemical potential is adjusted within the bulk gap in the trivial platform,  the in-gap states are localized magnetic states without a particle or hole counterpart. These states still exhibit the two artificial zero-energy crossings in the BdG Hamiltonian and the $\bZ_2$ index only affects the physical form of the in-gap states.   

To show that the origin of the two zero-energy crossings stems from the two orbitals, let's consider the limit where the exchange coupling $M_z$ from a single magnetic impurity goes to $+\infty$. In this limit, the total angular momentum ($J_z$) in the $z$ direction must be reduced by $2\times \hbar/2$ due to the spin flip in the two orbitals, compared to the system without the magnetic impurity. As discussed in the supplementary material~\cite{supplement_YSR}, each zero-energy crossing leads to a $-\hbar/2$ change in the angular momentum. Thus, there must be two zero-energy crossings. 

%
%
\begin{figure}[!t]
  \centering
\includegraphics[width=0.49\textwidth]{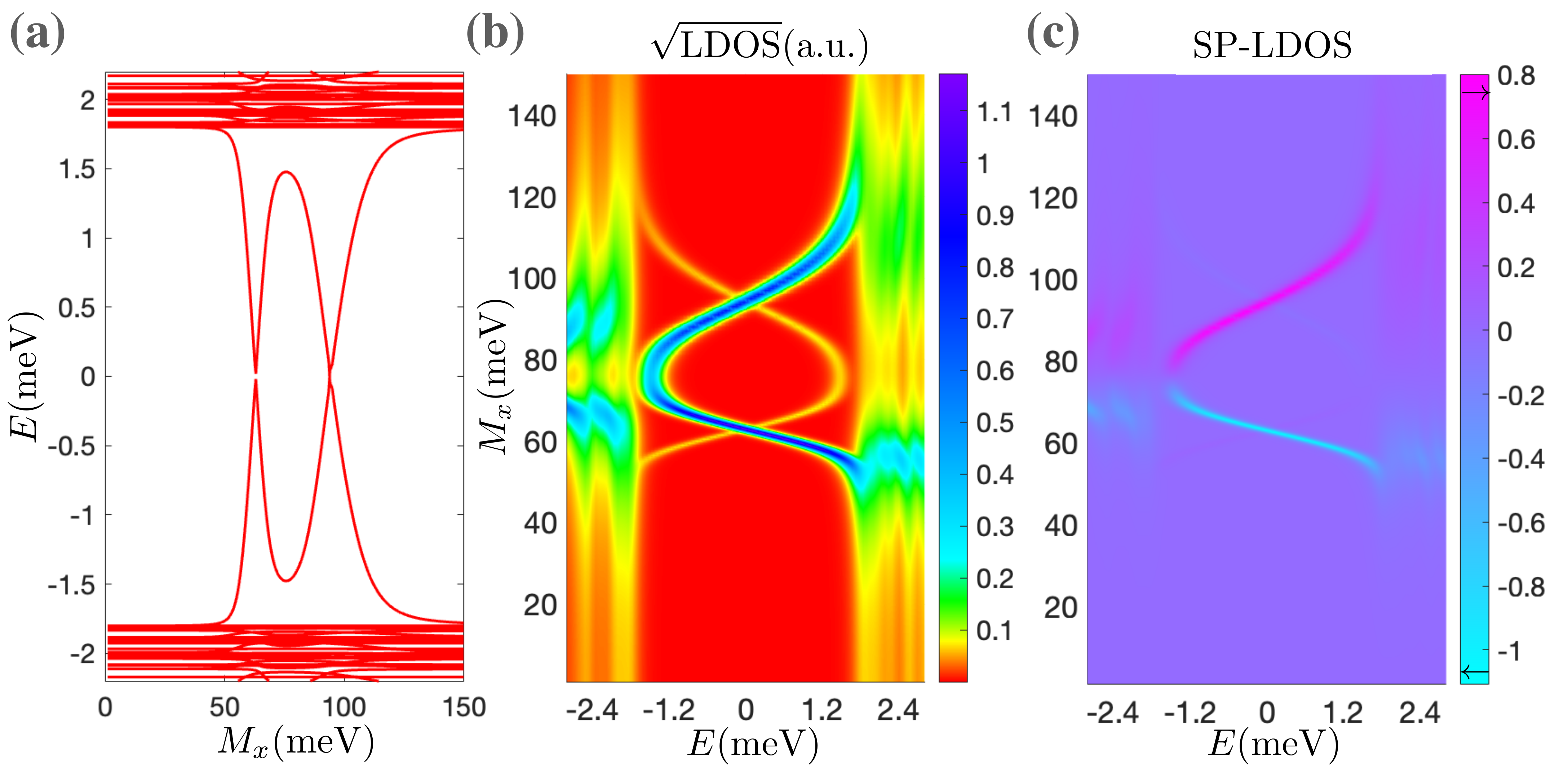}
  \caption{Evolution of the YSR states with {\it in-plane} exchange coupling $M_x$ ($\mu=5$meV, $m=2$, $M_y=M_z=0$). (a) Between the two zero-energy crossings, the YSR states evolve inside the SC gap with increasing $M_x$. (b) LDOS at the magnetic impurity showing the spectral weight of the YSR states. A square root scale bar is used to reveal the small particle-hole counterpart of the peak. (c) SP-LDOS showing the YSR state changes from spin-left ($-x$) to spin-right ($+x$) polarization after bending back.   }\label{Mx}
\end{figure}

We have studied the physics of the YSR states induced by a magnetic impurity in $s$-wave superconductors with topological nontrivial $\bZ_2$ bands. Our findings of two YSR states that evolve with exchange coupling and exhibit two zero-energy crossings are consistent with the experimental observations at the Fe adatoms deposited on the surface of ${\rm FeTe}_{0.55}{\rm Se}_{0.45}$ superconductors~\cite{Fan:2021td}. A closer comparison reveals that after the first zero-energy crossing, in the experiment~\cite{Fan:2021td}  the energy of the YSR state remains in the SC gap away from the continuum and forms the second crossing, which deviates from  the behavior shown in Fig.~\ref{m2}(a), where the inset displays the two YSR states crossing each other without hybridization at the gap edge. We argue that the absence of the YSR crossing is due to the canting of the magnetic moment of the Fe adatom away from the $z$-direction, which induces an important in-plane component of the exchange coupling ($M_x,M_y$).
%
In Fig.~\ref{Mx}, we show the evolution of the YSR states induced by the exchange coupling $M_x\tau_z\rho_0\sigma_x$ along the $x$ direction, while other parameters are the same as in Fig.~\ref{m2}. Fig.~\ref{Mx}(a,b) shows that after the first crossing, the energy of the YSR state stays inside the SC gap and forms the second crossing, as the exchange coupling increases. Thus, the exchange coupling in the in-plane direction brings consistency with the experiments. The reason is that the in-plane exchange coupling $M_x$ and finite chemical potential $\mu\neq0$~\cite{supplement_YSR} lead to the hybridization of the two YSR states so that the crossing point at the finite energy in the inset in Fig.~\ref{m2}(a)  is gapped and the new YRS state bend back to form the second zero-energy crossing. The agreement further supports the experimental observed evolution of the in-gap states as the novel YSR states in a superconductor with a topological nontrivial $\bZ_2$ band structure, and provide insights into their transition to the quantum anomalous vortex core states~\cite{PhysRevX.9.011033} near Fe adatoms in ${\rm FeTe}_{0.55}{\rm Se}_{0.45}$~\cite{Fan:2021td}.

We considered only the already challenging vortex-free solution and found that the YSR bound states generically exhibit two zero-energy crossings with increasing exchange coupling. The predicted second crossing at a larger exchange coupling is not seen experimentally~\cite{Fan:2021td}, but replaced instead by the coalescing of YSR states at zero-energy and the pinning of the zero-energy bound states with further increasing of the exchange coupling. This discrepancy indicates that increasing the normal state conductance, the experiments observed a transition out of the vortex-free YSR states, and mostly likely into a quantum anomalous vortex state~\cite{PhysRevX.9.011033} hosting a Majorana zero mode.

Our findings also shed light on the experimental detection of the YSR states using spin-polarized STM.
In the experiment using spin-polarized STM, but without pushing the tip closer to the Fe adatoms~\cite{PhysRevLett.126.076802}, the exchange coupling of the Fe magnetic moment may not be strong enough to cause the YSR state to pass through the first crossing. Therefore, the highest peak of the LDOS inside the SC gap has spin-down polarization and appears at a positive energy, while the second-highest peak with spin-up polarization is located at the opposite negative energy. In reality, the STM tip is not 100$\%$ polarized (the polarization is presumably less 10$\%$). Even if the LDOS peaks are $100\%$ polarized, the spin-polarized STM tip may pick up only a small difference between the two spins directions, which may be responsible for the current experimental observation. Our findings suggest that the spin-polarized STM tip can be used to distinguish the two energy crossings as the tip distance is reduced. As shown in Fig.~\ref{m2}(c), the main peak near the first crossing is spin-down polarized, while the it becomes spin-up polarized near the second crossing. The spin check in the future can confirm the monotonically-increasing exchange coupling with the decreasing tip distance~\cite{Fan:2021td}. 
 Finally, the recent ultra-high-resolution STM experiments show multiple YSR states inside the SC gap of ${\rm FeTe}_{0.55}{\rm Se}_{0.45}$~\cite{Machida_private}. Our results hint at the emergence of multiple YSR states from additional bulk bands derived from other atomic orbitals. 

We would like to thank Tadashi Machida, Kun Jiang, Shuheng Pan, and Dongfei Wang  for insightful discussions. Z.W. acknowledges the support of the U.S. DOE, Basic Energy Sciences Grant No. DE-FG02-99ER45747.

\bibliographystyle{apsrev4-1}
\bibliography{SC_v2}

%

 \clearpage
\newpage

\onecolumngrid
\widetext

\begin{center}
\textbf{
\large{Supplementary Materials for}}
\vspace{0.4cm}

\textbf{
\large{
``Yu-Shiba-Rusinov states on a superconductor with topological $\bZ_2$ bands" }
}
\end{center}

\setcounter{secnumdepth}{3}
\setcounter{equation}{0}
\setcounter{figure}{0}
\renewcommand{\theequation}{S-\arabic{equation}}
\renewcommand{\thefigure}{S\arabic{figure}}
\renewcommand\figurename{Supplementary Figure}
\renewcommand\tablename{Supplementary Table}
\newcommand\Scite[1]{[S\citealp{#1}]}

\vspace{0.1cm}

\begin{center}
\textbf{Authors:}
Ching-Kai Chiu and Ziqiang Wang
\end{center}

\vspace{0.5cm}


This supplementary information is organized as follows. In Sec.~\ref{I}, we discuss additional properties of YSR states at zero-energy crossings. In Sec.~\ref{III}, we derive analytically the two zero-energy crossings of the YSR states stemming from the SC surface and the bulk states. In Sec.~\ref{I2}, the evolution of the YSR states as a function of the exchange coupling is obtained numerically for a system with the Fermi level cutting through both the bulk and surface bands.
In Sec.~\ref{IV}, we study both numerically and analytically the magnetic impurity induced in-gap states of our model in the $\bZ_2$ trivial region. In Sec.~\ref{in-plane}, we show that the absence of the YSR energy crossings at finite energy stems from non-zero in-plane exchange coupling and finite chemical potential.  






\section{Additional properties of YSR states at zero-energy crossings}

	Studying physical properties of YSR states at zero-energy crossings are important for understanding those low energy states. Here we discuss some additional properties --- 1. total angular momentum number and fermion parity, 2. spatial distribution of LDOS, 3. different exchange coupling strengths in different orbitals.

\subsection{The change of total angular quantum number and fermion parity } \label{I}

\begin{figure}[!b]
  \centering
\includegraphics[width=0.49\textwidth]{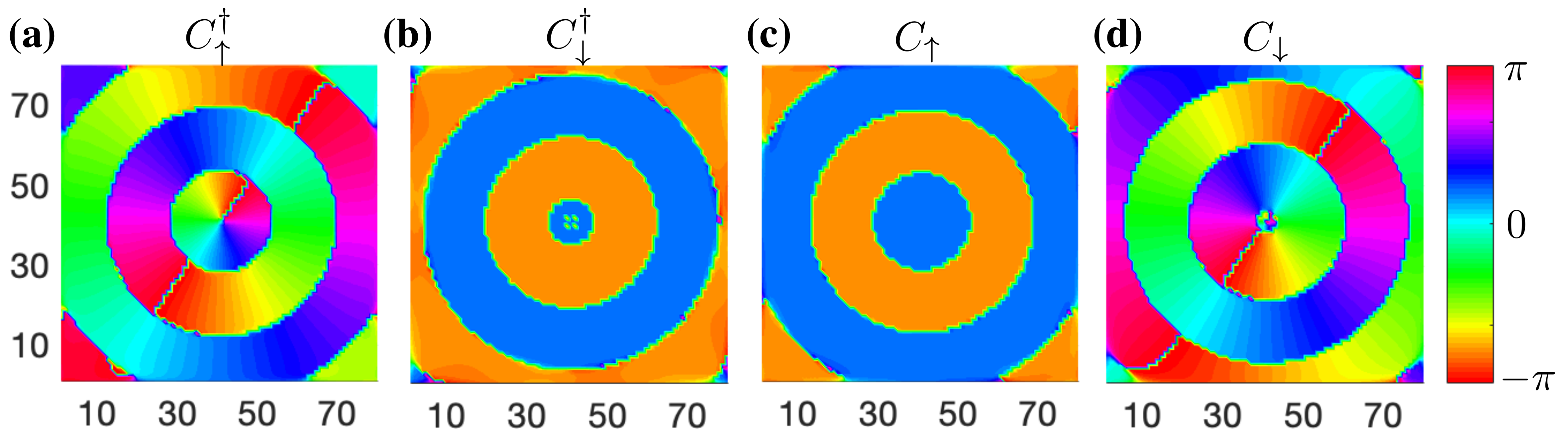}
  \caption{Spatial distribution (in unit of nm) of the wavefunction phases of the in-gap state at the positive energy before the first zero-energy crossing in Fig.~\ref{m2}(a) ($\mu=5$meV, $m=2$, $M_z<M_z^{\rm{c1}}=61$meV), showing a phase winding for spin-up particle and spin-down hole, while  no winding for spin-down particle and spin-up hole.
  The boundary of the $\pi$-phase difference indicates the node of the wavefunction. The four panels are obtained for the first orbital, with the second orbital having the same phase windings.    }\label{particle hole phase}
\end{figure}

The zero-energy crossings manifest changes of quantum numbers. We discuss changes of total angular momentum and fermion parity in this section. In the presence of the spin-orbital coupling, spin $\sigma_z$ is not a good quantum number anymore. Instead, the changes in the total angular quantum number with its projection on the $z-$axis ($J_z$) should be considered. Denoting the eigenstates
$\Psi_j(\phi,r)=(e^{iN_{1j}\phi}\bu_{\uparrow j}(r),e^{iN_{2j}\phi}\bu_{\downarrow j}(r),e^{iN_{3j}\phi}\bv_{\uparrow j}(r),e^{iN_{4j}\phi}\bv_{\downarrow j}(r))$, where
$N_{1j,\dots,4j}$ are the orbital angular quantum numbers governing the phase winding of the wavefuctions around the impurity center, the orbital angular momentum and the spin in the $z$-direction are given by~\cite{PhysRevB.93.174505}
\begin{align}
L_z =& \hbar \sum_{E_j>0} \int d^3 \bm{r} \big [ N_{3j}|\bv_{\uparrow j}(r)|^2 + N_{4j} |\bv_{\downarrow j}(r)|^2 \big ], \\
S_z =& \frac{\hbar}{2} \sum_{E_j>0} \int d^3 \bm{r} \big [ |\bv_{\uparrow j}(r)|^2 - |\bv_{\downarrow j}(r)|^2 \big ].
\end{align}

We are interested in the changes of the quantum numbers as the exchange coupling $M_z$ passes through the first zero-energy crossing at $M_z^{c1}$. Before the crossing, the spatial distribution of the wavefunction phase of the in-gap state at positive energy $E_0>0$ is shown in Fig.~\ref{particle hole phase}. The wavefunctions of the spin-up particle and spin-down hole exhibit the counterclockwise phase winding centered at the Fe adatom.
The wavefunction thus has the form $\Psi_0^{\rm{be}+}(\phi,r)=(e^{i\phi}\bu_{\uparrow0}(r),\bu_{\downarrow0}(r),
\bv_{\uparrow0}(r),e^{i\phi}\bv_{\downarrow0}(r))$. Likewise, due to the particle-hole symmetry, the wavefunction of the particle-hole partner at energy $-E_0$ is given by $\Psi_0^{\rm{be}-}(\phi,r)=(\bv_{\uparrow0}^*(r),e^{-i\phi}\bv_{\downarrow0}^*(r),
e^{-i\phi}\bu_{\uparrow0}^*(r),\bu_{\downarrow0}^*(r))$. 
Since the wavefunction of the lowest positive energy state switches from $\Psi_0^{\rm{be}+}(\phi,r)$ to $\Psi_0^{\rm{af}+}(\phi,r)=\Psi_0^{\rm{be}-}(\phi,r)$ and the rest of the states with $j\neq0$ are unchanged after the zero-energy crossing, due to the particle-hole symmetry of the BdG Hamiltonian, the differences in $L_z$ and $S_z$ are given by
\begin{align}
\Delta L_z = & -\hbar \int d^3 \bm{r}  \big [ |\bu_{\uparrow0}(r)|^2 + |\bv_{\downarrow0}(r)|^2 \big ] \\
\Delta S_z = &  \frac{\hbar}{2}  \int d^3 \bm{r}  \big [ |\bu_{\uparrow0}(r)|^2 -  |\bu_{\downarrow0}(r)|^2 - (\bu\to \bv)\big],
\end{align}
leading to the change in the total angular quantum number in the $z$ direction 
\be
\Delta J_z = \Delta L_z + \Delta S_z = - \frac{\hbar}{2}.
\ee
As a result, the zero-energy crossing leads to $-\hbar/2$ angular momentum change in the entire BCS wavefunction. Similarly, the second zero-energy crossing also leads to a $-\hbar/2$ change in $\Delta J_z$.

\subsection{The LDOS spatial line cuts for the two crossing points}

\begin{figure}[!tbp]
  \centering
\includegraphics[width=0.55\textwidth]{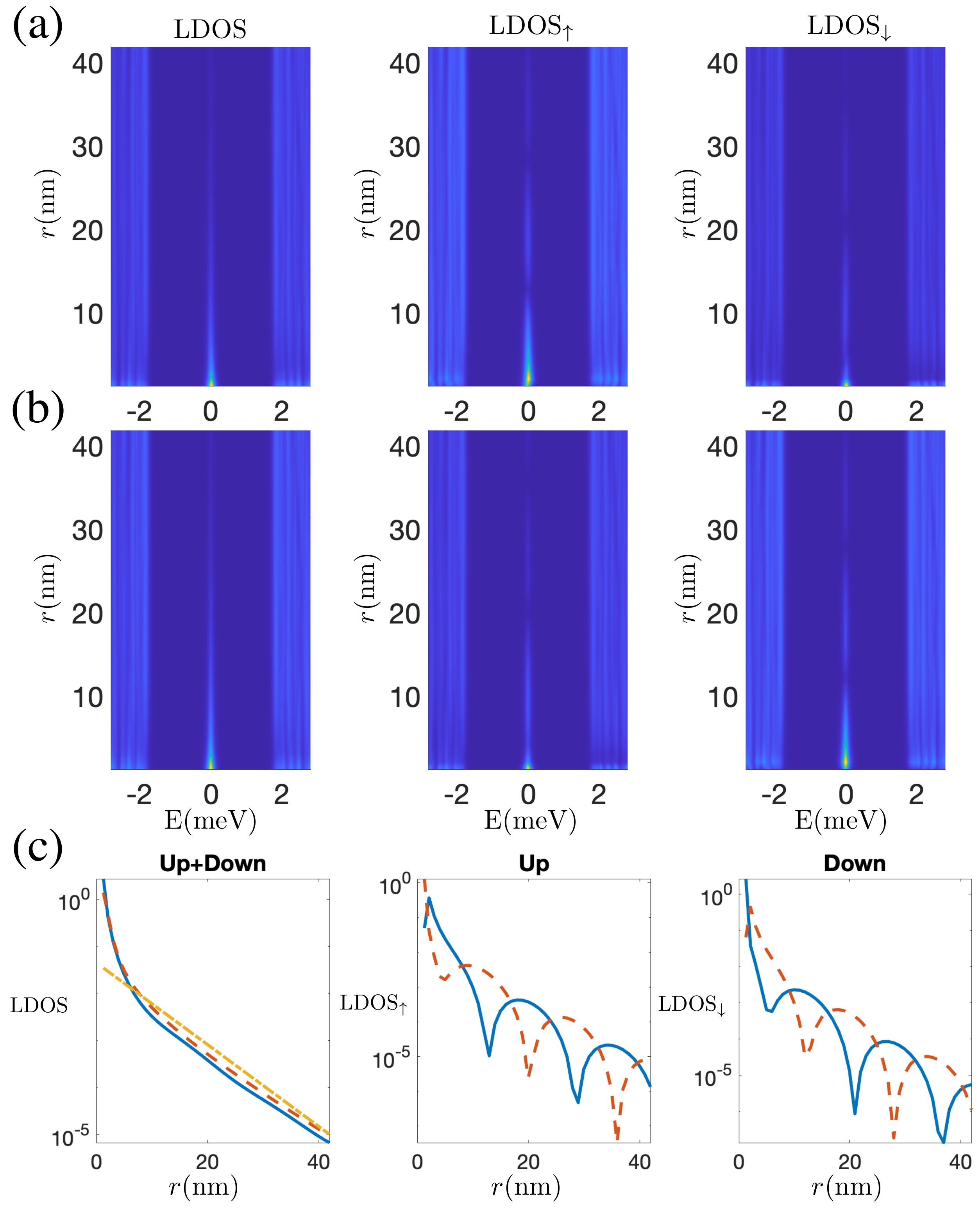}
  \caption{The LDOS line cut from the impurity center ($r=0$) at (a) the first crossing (b) the second crossing can be distinguished by spin LDOS. (c) shows the LDOS of the zero-bias peaks move away from the impurity center for the first (blue solid lines) and second (red dashed lines) crossings. The dashed dotted orange line indicates the fitting of the LDOS exponential decay and the decay length is approximately $5$nm.     }\label{crossing_line_cut}
\end{figure}

To compare the feature of the two zero-energy crossings, we plot Fig.~\ref{crossing_line_cut} showing the spatial evolution of the local density of states (LDOS) spectra along a line cut across the impurity calculated for the two exchange couplings corresponding to the two zero-energy crossings ((a) and (b)). Clearly, the total LDOS (left panel in (a) and (b)) cannot distinguish between the two zero-energy crossings. However, the spin-resolved LDOS (middle and right panels in (a) and (b)) shows marked difference at the two zero-energy crossings. As discussed in the main text, at the impurity center ($r=0$), spin-down dominates at the first crossing while spin-up dominates the LDOS at the second crossing. We thus predict that spin-resolved STM can probe spin-polarized LDOS at the impurity and distinguish between the first and second zero-energy crossings. The spectral peak at zero-energy decays exponentially when moving away from the impurity as shown in panel (c). The decay length is around $5$nm, which is shorter than the coherence length ($\nu_F/\Delta_0=13.9$nm)~\cite{Zhang2018Observation,Wang2018Evidence,Chiueaay0443}. The spin texture of the zero-bias-peak line cut exhibits spatial oscillation with periodicity close to the Fermi wave length of the Dirac cone ($\pi/k_{F}=15.7$nm)~\cite{Zhang2018Observation,Wang2018Evidence,Chiueaay0443}.

\subsection{Different exchange coupling for different orbitals}

\begin{figure}[!bp]
  \centering
\includegraphics[width=0.45\textwidth]{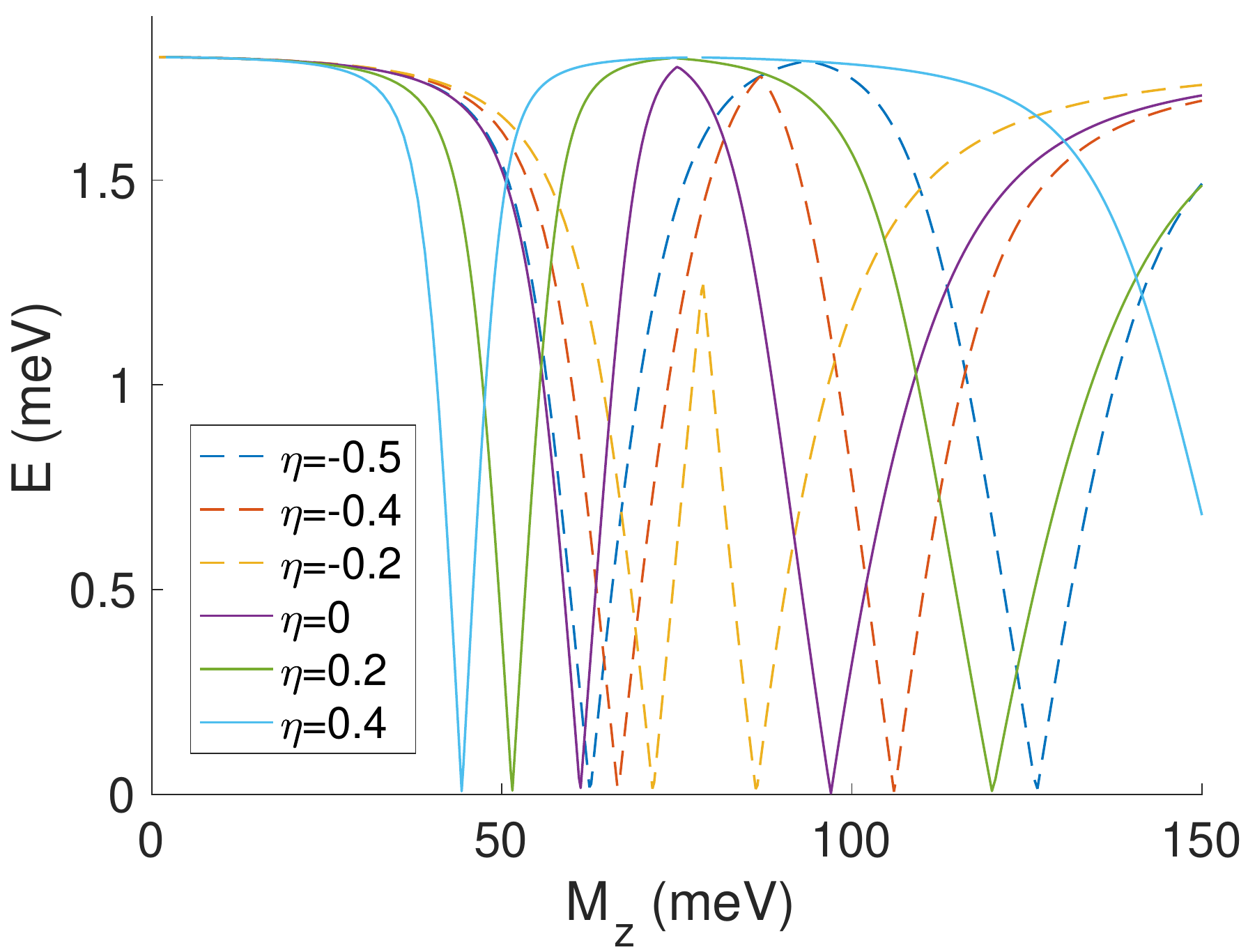}
  \caption{The points of the two zero-energy crossing varies as the different orbital strength $\eta$ of the exchange coupling is changed.  }\label{different_orbitals}
\end{figure}

In reality, different orbitals might exhibit different strengths of exchange coupling. Hence, we consider a form of the exchange coupling $M_z\tau_z (\rho_0 + \eta \rho_z)\sigma_z$, where $\eta$ adjusts the difference of the orbitals; the average exchange coupling is still $M_z$. Fig.~\ref{different_orbitals} shows the double-crossing is always present and at $\eta=-0.2$ the two zero-energy crossing points are closest. As $\eta$ moves away from $-0.2$, the distance between the two crossing points is increased.

\section{Analytical solution of the YSR states in a superconductor with $\bZ_2$ bands  } \label{III}

We demonstrate an approach to find the condition of the zero-energy YSR state by solving $H_+$(\ref{Hpm}) in the section. Our focus is on the low-energy physics near $\bar{\Gamma}$ ($k_x=k_y=0$) so that we transform the lattice model to the continuous model in the two in-plane directions by performing approximations $\sin k_x \rightarrow k_x,\ \sin k_y \rightarrow k_y$ and $\cos k_x, \cos k_y \rightarrow 1$.
To study YSR states on the surface, we transfer $H^+$ in the second quantization expression of the momentum space in the $z$ direction to the real space
\begin{small}
\begin{align}
\hat{H}^+=& v_F \sum_{-L_z < z \le 0} \Big \{ C_z^\dagger (\frac{\Delta_0}{v_F} \rho_z \sigma_0 - D \rho_y \sigma_y + k_x \rho_0 \sigma_x - k_y \rho_z \sigma_y) C_z  \nonumber \\
&+ \big [  \frac{t_0}{2v_F}C^\dagger_{z} (\rho_y \sigma_y - i\rho_x \sigma_y ) c_{z-1}+h.c.\big] \Big \} -M_z\delta(\vec{r})C^\dagger_0 \rho_z \sigma_z C_0,
\end{align}
\end{small}
where $D=-m+2$ and $t_0$ is the strength of the nearest neighbor hopping ($t_0=v_F$ in our case). Since the in-gap YSR states are always localized on the surface, we express the block Hamiltonian from the (001) surface
\be
H^+= v_F
\left(
\begin{array}{ccccc}
 H_1-m_z\delta(\vec{r})\sigma_z & i D \sigma_y & 0_{2} & 0_{2} & \cdots \\
 -i D \sigma_y &  H_2+m_z\delta(\vec{r})\sigma_z  & it\sigma_y & 0_{2\times 2} & \cdots \\
 0_{2} & -it\sigma_y &  H_1 & i D \sigma_y  & \cdots \\
 0_{2} &  0_{2} & -i D \sigma_y  &  H_2  &   \cdots    \\
\vdots & \vdots & \vdots & \vdots  &  \ddots  \\
\end{array}
\right)
\ee
where $H_{\alpha}=k_x\sigma_x +(-1)^{\alpha} k_y \sigma_y + \Delta \sigma_z$ ($\alpha=1,2$), $m_z=M_z/v_F$, $t=t_0/v_F$, $\Delta = \Delta_0/v_F$, and $0_2$ represents the $2\times 2 $ null matrix. We note that the the first $4\times 4$ block represents the (001) surface. To interpret this expression, we start with $D=0 (m=2)$. The first $2\times 2$ block, which is isolated, indicates the SC surface Dirac cone, while the second $4\times 4 $ block including the the exchange coupling manifests a part of the bulk bands. As $D$ moves away from $0$, those blocks are not isolated. Since the YSR is localized on the surface, we simplify our problem to the first $6\times 6$ block. Although the first $6\times 6$ block couples to other layers as $D\neq 0$, this is a good approximation for small $D$.
Now we solve the Schr\"{o}dinger equation for the zero-energy YSR state
\be
E\Psi(\vec{r})= (H_0 - m_z \delta (\vec{r})Y)\Psi(\vec{r}),
\ee
where
\be
H_0(\vec{\hat{k}})\equiv
\left(
\begin{array}{ccccccccc}
 \Delta -M_z & \hat{k}_+ & 0 & D & 0 & 0  \\
\hat{k}_- & -\Delta +M_z  & -D & 0 & 0 & 0  \\
 0 & -D & \Delta +M_z  & \hat{k}_- & 0 & +t  \\
 D & 0 & \hat{k}_+ & -\Delta -M_z &   -t & 0 \\
0 & 0 & 0 & -t &  \Delta & \hat{k}_+ & \\
0 & 0 & +t & 0 & \hat{k}_- & -\Delta  \\
\end{array}
\right), Y\equiv  \left(
\begin{array}{ccccccccc}
1 & 0 & 0 & 0 & 0 & 0  \\
0 & -1  & 0 & 0 & 0 & 0  \\
 0 & 0 & -1  & 0 & 0 & 0  \\
0 & 0 & 0 & 1 &   0 & 0 \\
0 & 0 & 0 & 0 &  0 & 0  \\
0 & 0 & 0 & 0 & 0 & 0  \\
\end{array}
\right),
\ee
where momentum operators are $k_\pm=\hat{k}_x\pm i\hat{k}_y$.
We rewrite the wavefunction on the L.H.S. in the momentum form $\Psi(\vec{r})=\int \frac{d^2\vec{k}}{(2\pi)^2}e^{i\vec{k}\cdot \vec{r}} \Psi_{\vec{k}}$ and rearrange the equation
\be
\int \frac{d^2\vec{k}}{(2\pi)^2} e^{i\vec{k}\cdot \vec{r}} (E - H_0(\vec{k}))\Psi_{\vec{k}} = - m_z \delta (\vec{r})Y \Psi(\vec{r})
\ee
By multiplying $ \int d^2\vec{r} e^{-i\vec{k}\cdot \vec{r}}$, the Hamiltonian is written in the form of
\be
(E - H_0(\vec{k}))\Psi_{\vec{k}} = - m_z Y \Psi(0)
\ee
We particularly search for the YSR states with zero energy ($E=0$); by multiplying $-H_0(\vec{k})^{-1}$, the equation is given by
\be
\Psi_{\vec{k}} =  H_0(\vec{k})^{-1}m_z Y \Psi(0)
\ee
By performing the integration $\int \frac{d^2\vec{k}}{(2\pi)^2} $, we have
\be
\Psi(0)=  m_z \int \frac{d^2\vec{k}}{(2\pi)^2}  H_0(\vec{k})^{-1} Y \Psi(0)
\ee
Hence, the wavefunction has solutions when
\be
\det \big(\bI_{6\times 6} - m_z \int_0^\Lambda H_0(\vec{k})^{-1} Y  \frac{d^2\vec{k}}{(2\pi)^2}  \big) =0
\ee
where $\Lambda$ is the momentum cut off in the lattice system. Due to the special form of $Y$, the determinant of the $6\times6$ matrix equals to the determinant of the first $4\times 4$ block. Furthermore, after the integration, the odd functions of $k_x$ and $k_y$ vanish. Hence, we can rewrite the determinant in a simpler and explicit form
\be
\det \big(\bI_{6\times 6} - m_z \int_0^\Lambda H_0(\vec{k})^{-1} Y  \frac{d^2\vec{k}}{(2\pi)^2}  \big)=\det \big(\bI_{4\times 4} - \frac{m_z}{2\pi} \int_0^\Lambda {H'_0}^{-1}(\vec{k})Y' kdk \big) =0
\ee
where
\begin{align}
{H'_0}^{-1}
=
\bma
\Delta \alpha & 0 & 0 & D\beta \\
0 & -\Delta \alpha & - D \beta & 0 \\
0 & - D \beta & \Delta \beta & 0 \\
D\beta & 0 & 0 & -\Delta \beta \\
\ema
,\quad
Y'=\bma
1 & 0 & 0 & 0 \\
0 & -1 & 0 & 0 \\
0 & 0 & -1 & 0 \\
0 & 0 & 0 & 1 \\
\ema, \\
\alpha = \frac{k^2 + \Delta^2 + t^2}{(k^2+\Delta^2)(k^2+ \Delta^2 + t^2 + D^2)},\quad \beta = \frac{1}{k^2+ \Delta^2 + t^2 + D^2}
\end{align}
After solving the integration, we have
\be
0= \big[ -1 + m_z (a-b)\Delta+m_z^2B(a^2 \Delta^2+D^2b^2)   \big]^2,
\ee
where
\begin{align}
a=&\frac{\Delta }{4\pi (D^2+t^2)} \Big \{ t^2 \ln (1+\frac{\Lambda^2}{\Delta^2})+D^2 \ln (1+\frac{\Lambda^2}{\Delta^2+ t^2+D^2}) \Big \}, \\
b=&\frac{\Delta }{4\pi } \ln (1+\frac{\Lambda^2}{\Delta^2+t^2+D^2}),
\end{align}
The localized states have zero energy when the normalized exchange coupling obeys 
\be
m_{z\alpha}^+ =\frac{2}{a-b - (-1)^{\alpha} \sqrt{(a+b)^2+ 4b^2D^2/\Delta^2}},\quad \alpha=1,2
\ee
Using the effective time reversal symmetry (\ref{time reversal symmetry}), we have the zero energy YSR states for $H_-$ when $m_{z\pm}^-=-m_{z\pm}^+$. We recover the normalized parameters $(m_z,\Delta,t,D)$ to the physical parameters in the main text $(M_z,\Delta_0,v_F,D)$, the relation between the zero energy and the exchange coupling is written as
 \be
M_{z\alpha}^+ =\frac{2v_F^2}{A-B \pm \sqrt{(A+B)^2+ 4B^2D^2v_F^2/\Delta^2_0}},\quad \alpha=1,2 \label{zero solution}
\ee
where 
\begin{small}
\begin{align}
A=&\frac{\Delta_0 }{4\pi (D^2+1)} \Big \{  \ln (1+\frac{\Lambda^2 v_F^2}{\Delta^2_0})+D^2 \ln (1+\frac{\Lambda^2v_F^2}{\Delta_0^2+ v_F^2(D^2+1)}) \Big \}, \\
B=&\frac{\Delta_0 }{4\pi } \ln (1+\frac{\Lambda^2v_F^2}{\Delta^2_0+v_F^2(D^2+1)}).
\end{align}
\end{small}
Using the effective time reversal symmetry (\ref{time reversal symmetry}), we have the zero energy YSR states for $H_-$ when $M_{z\pm}^-=-M_{z\pm}^+$. Consider $m=2 (D=0)$ as the surface Dirac node is completely localized on the first layer. The relation between the zero energies and the exchange coupling can be written in a simple form in Eq.~\ref{m2simple} in the main text.

\section{The case with Fermi level inside the bulk band} \label{I2}

\begin{figure}[!t]
  \centering
\includegraphics[width=0.69\textwidth]{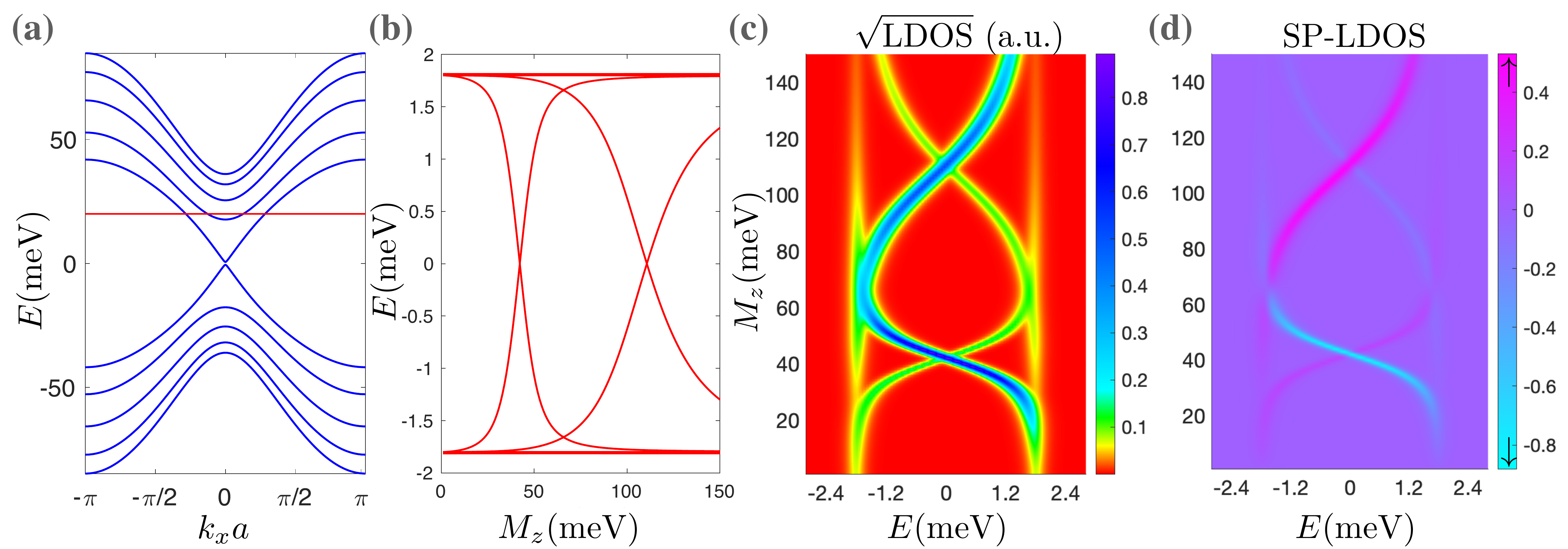}
  \caption{ The Fermi level cuts through the bulk band ($\mu=20$meV, $m=2$, $M_x=M_y=0$). (a) The semi-surface band in the spectrum of $H_{\rm{TI}}$ in eq.~1 at $k_y=0$. Due to the finite size effect ($L_z=5$), the semi-surface band from the surface Dirac cone does not exhibit a clear localization. (b,c) The spectrum and LDOS of the BdG Hamiltonian show two zero-energy crossing as $M_z$ increases. (d) The spin polarization is similar to the Fermi level in the gap (Fig.~2(c)).  }\label{cp20}
\end{figure}

For a topological insulator with doping, the Fermi level can be moved to the bulk bands to induce superconductivity and simultaneously pass through the Dirac surface band. Particularly, Cu$_x$Bi$_2$Se$_3$~\cite{WrayCavaHasan,PhysRevLett.107.217001}, whose Fermi level cuts through the bulk and surface bands, is one of the material examples. To understand the YSR physics on its surface, by keeping the same parameters in Fig.~2, we change the chemical potential from $\mu=5$meV to $\mu=20$meV and the topological phase tuning parameter from $m=2$ to $m=2.5$. Due to the finite size effect, as shown in Fig.~\ref{cp20}(a) the Fermi level cuts through a bulk band and a semi-surface band, which has a long localization length. Likewise, as the exchange coupling increases, as shown in Fig.~\ref{cp20}(b,c,d) the YSR states exhibit two zero-energy crossings and share the same similar spin polarization with the cases of the Fermi level ($\mu=5,0$meV in Fig.~\ref{m2},\ref{cp0}) in the gap. We can still borrow the physical understanding from sec.~\ref{III}. That is, the first zero-energy crossing mainly stems from the surface band, while the part of the bulk band near the surface leads to the second crossing.

\section{Impurity states in the $\bZ_2$ trivial region} \label{IV}

\begin{figure}[!b]
  \centering
\includegraphics[width=0.49\textwidth]{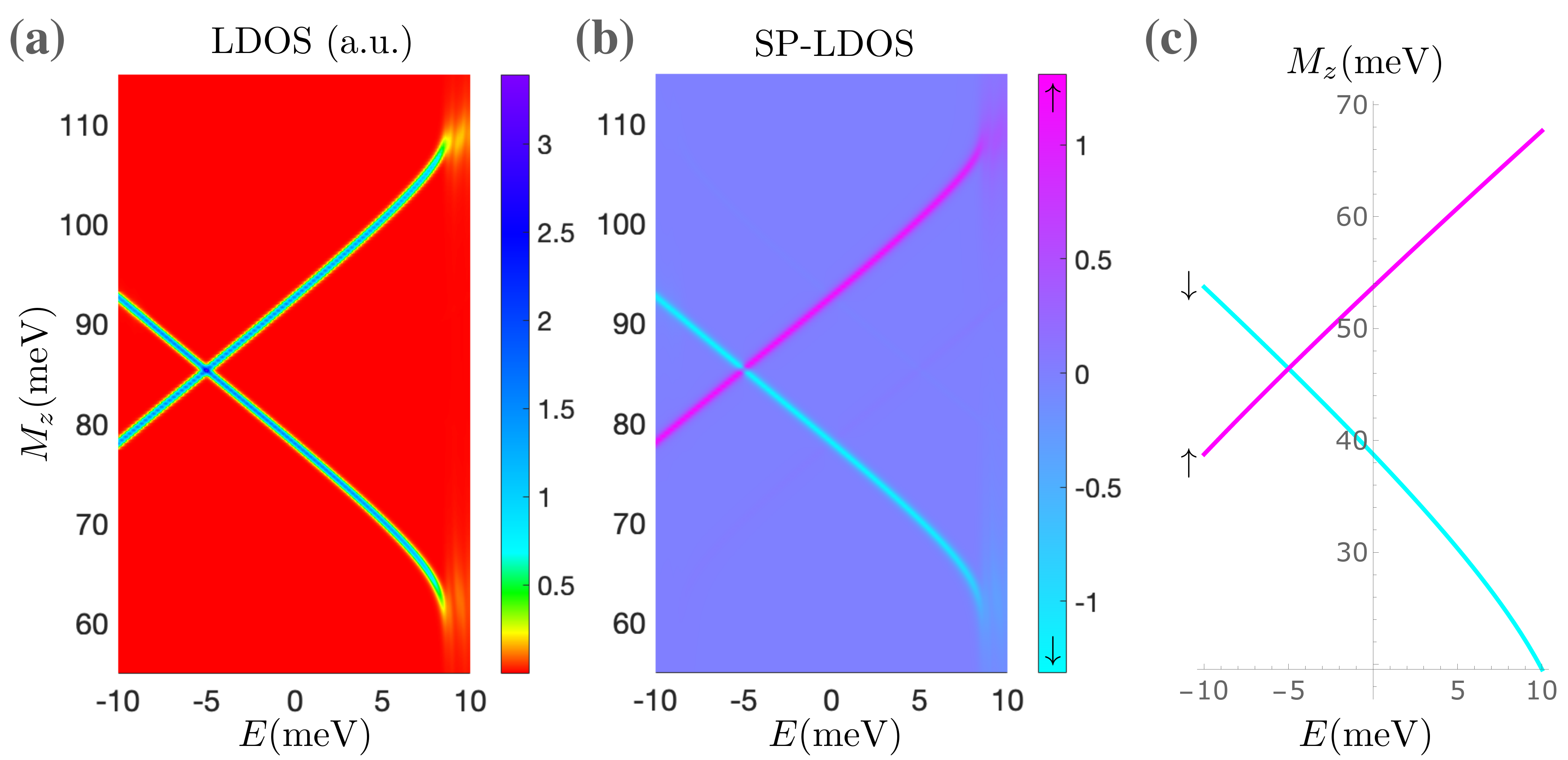}
  \caption{ The Fermi level is located in the trivial bulk gap ($\mu=5meV$, $m=3.3$, $M_x=M_y=0$). (a) As $M_z$ increases, the simulation shows that the first LDOS peak moves from the positive energy gap to the negative one. The second peak appears later and moves in the opposite direction. (b) the first peak is spin-down polarized, while the second peak is spin-up polarized. (c) The effective quadratic model (\ref{quadratic}) exhibits the consistent spin polarization order.  We adopt $\Lambda^2/2m=50$meV, $m=100/\pi^2$meV, $g_+=12$meV, and $g_-=22$meV.
  }\label{m33}
\end{figure}

	We study the evolution of the in-gap states in the absence of the surface Dirac cone to compare with the YSR states in the topological region. That is, the Hamiltonian $H_{\rm{TI}}$ (\ref{TIH}) is adjusted to the trivial region by choosing $m=3.3$. The spectrum in Fig.~\ref{schametics}(c) shows the bulk gap is close to 8meV due to the finite size effect, in comparison with the smallest bulk gap $|v_F(3-m)|=7.5$meV at $Z$ point in the momentum space. We note that in the absence of the gapless surface states the bulk gap is greater than the superconductor gap $\Delta_0=1.8$meV. The gap of the BdG Hamiltonian $H_{\rm{BdG}}$ mainly stems from the insulator gap so that the superconductivity is suppressed. In this regard, the in-gap states are just normal localized states exhibiting a single LDOS peak, and the spectral weights of their particle or hole counterparts completely vanish as shown in Fig.~\ref{m33}(a). This is different from the topological case with the superconducting Dirac cone surface where the LDOS peak from the in-gap YSR state is accompanied by a particle or hole non-vanishing counterpart (Fig.~\ref{m2}(b)). On the other hand, in this trivial case the spin-polarization in Fig.~\ref{m33}(b) is similar to the topological case (Fig.~\ref{m2}(c)); the first LDOS peak is spin-down polarized and moves to negative energy with increasing $M_z$, while the second one is spin up and moves to the opposite direction. Back to the current experiments~\cite{Fan:2021td,PhysRevLett.121.196803}, the in-gap state always exhibits two clear LDOS peaks at positive and negative energies; thus, we can rule out normal states from non-superconductivity.

	The in-gap states can be understood by solving the two types of the 2D quadratic Schr\"{o}dinger equations in the absence of superconductivity. In the presence of the single magnetic impurity, we use two Hamiltonians $h_\downarrow=(\vec{k}^2/2m+g_+)\sigma_0 + M_z \delta(\vec{r}) \sigma_z $ and  $h_\uparrow=-(\vec{k}^2/2m+g_-)\sigma_0 + M_z \delta(\vec{r}) \sigma_z $ to simplify the insulator model (\ref{TIH}), where $g_\pm$ represents the distance between the Fermi level and the quadratic energy dispersion. The effective Hamiltonian $h_\downarrow$ describes the insulator dispersion above the zero energy (Fig.~\ref{schametics}(c)), while $h_\uparrow$ describes the dispersion below the zero energy. Solving the eigenvalue problem~\cite{PhysRevB.88.155420} in the next paragraph, we have an in-gap spin-down state from $h_\downarrow$ and a spin-up one from $h_\uparrow$ (see the next section for the derivation). Furthermore, the relations between the exchange coupling and the energy are given by respectively
\begin{align}
M_z^{\downarrow\pm} =&\pm \frac{2\pi}{m\ln(1 + \frac{\Lambda^2}{2m(g_+-E)}) }, \nonumber \\
M_z^{\uparrow\pm} =& \pm \frac{2\pi}{m\ln(1 + \frac{\Lambda^2}{2m(g_-+E)}) }, \label{quadratic}
\end{align}
where $\Lambda$ is the momentum cut off. The analytic result as shown in Fig.~\ref{m33}(c) is consistent with the numerical one in panel (b). In other words, as $M_z$ increases, the dispersion above the Fermi level leads to the spin-down state with decreasing energy, and the dispersion below the Fermi level leads to the spin-up state with increasing energy. Importantly, as $\mu>0$ ($g_+<g_-$), the spin-down polarized LDOS peak appears before the spin-up polarized LDOS peak. 

We show the calculation for a localized in-gap state at a magnetic impurity below the quadratic dispersion. The Hamiltonian reads
\be
h_{\downarrow}= (\vec{k}^2/2m+g_+)\sigma_0 + M_z \delta(\vec{r}) \sigma_z ,
\ee
where $g_+$ is the distance of the Fermi level from the quadratic energy dispersion. Another in-gap state from the Hamiltonian
\be
h_\uparrow=-(\vec{k}^2/2m+g_-)\sigma_0 + M_z \delta(\vec{r}) \sigma_z,  \label{hdown}
\ee
can be solved by a similar approach in the following. We write the Hamiltonian $h_\downarrow $ in the form of the Sch\"{o}dinger equation
\be
(E-H_0)\psi(\vec{r})=  M_z \delta(\vec{r}) \sigma_z  \psi(\vec{r}),
\ee	
where $H_0\equiv\vec{\hat{k}}^2/2m+g_+$ and $\vec{\hat{k}}$ is a momentum operator in the real space. By dividing the equation by $(E-H_0)$ and multiplying $ \int \frac{d^2 \vec{k}}{(2\pi)^2} e^{i\vec{k}\cdot \vec{r}}$ we transform the equation to the momentum space
\be
\psi_{\vec{k}}= M_z(E-H_0)^{-1}\sigma_z \psi(0)
\ee
Multiplying $ \int \frac{d^2\vec{k}}{(2\pi)^2}$, we obtain
\be
\psi(0)=\frac{M_z}{(2\pi)^2}\int^\Lambda_0 \frac{d^2\vec{k}}{E-H_0}\sigma_z \psi (0),
\ee
where $\Lambda$ is the momentum cut off. To have a solution of $\psi(0)$, the determinant must vanish.
\be
\det \big( \sigma_0 + \frac{M_z m}{2\pi}\ln (1 + \frac{\Lambda^2}{2m(g_+-E)}) \sigma_z \big)=0
\ee	
We have the relation between $M_z$ and $E$
\be
M_z^{\downarrow\pm} =\pm \frac{2\pi}{m\ln(1 + \frac{\Lambda^2}{2m(g_+-E)}) }.
\ee	
For $M_z^{\downarrow+}$, the wavefunction in the impurity is purely spin down $(0,1)^T$, while for $M_z^{\downarrow-}$, the wavefunction in the impurity is purely spin up $(1,0)^T$. Importantly, this spin-polarized wavefunction is a normal state without any particle or hole counterpart. Following the same recipe, we have another localized in-gap state obeying
\be
M_z^{\uparrow \pm} =\pm \frac{2\pi}{m\ln(1 + \frac{\Lambda^2}{2m(g_-+E)}) }.
\ee	
Likewise, for $M_z^{\uparrow+}$, the wavefunction in the impurity is purely spin up $(1,0)^T$, while for $M_z^{\uparrow-}$, the wavefunction in the impurity is purely spin down $(0,1)^T$. In our trivial case, with $\mu=5$meV the Fermi  level is close to the upper bands, so we choose $g_+=12$meV and $g_-=22$meV. Since Fig.~1b shows that the energy difference at $k_xa=0, \pi$ is $50$meV, the momentum cut off can be chosen to be $\pi$ and satisfies $\Lambda^2/2m\sim50$meV. With the values of the parameters, based on the $M_z-E$ relations above, Fig.~\ref{m33}(c) shows with the increasing $M_z$ the movements and the spin-polarization of the LDOS peaks are consistent with the numerical simulation in panel (a,b).

\section{The absence of the energy crossing at finite energy between the two zero-energy crossings} \label{in-plane}

\begin{figure}[!tbp]
  \centering
\includegraphics[width=0.65\textwidth]{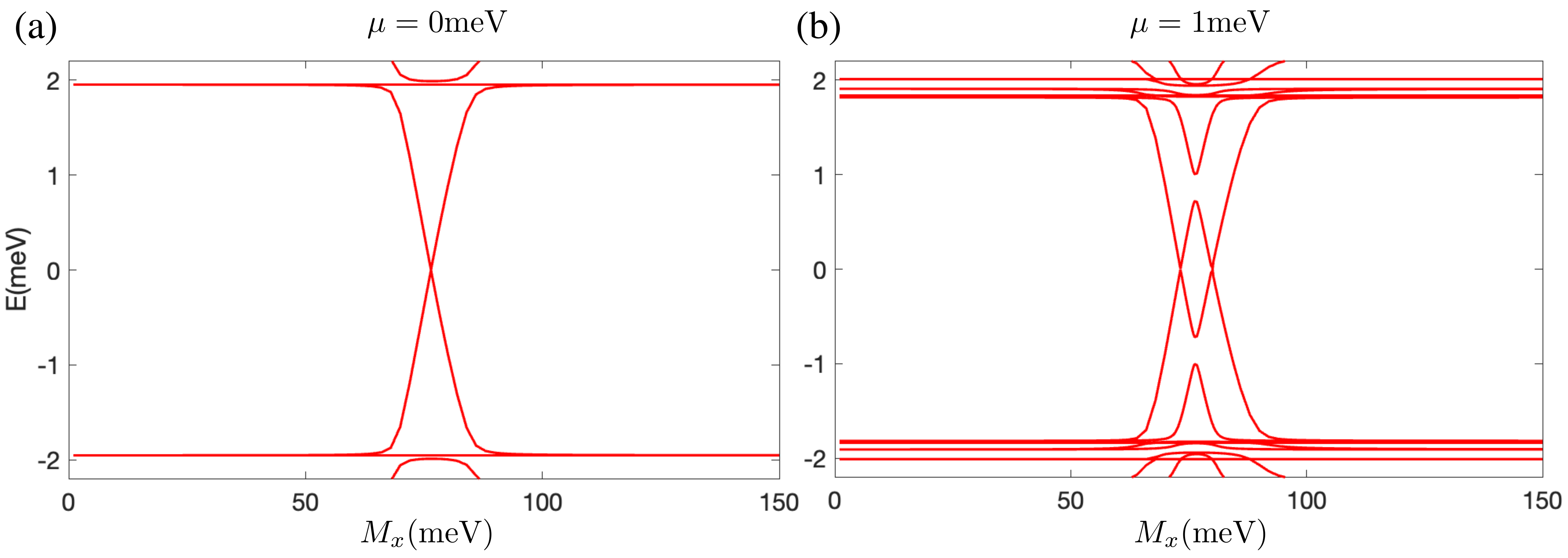}
  \caption{(a) When the chemical potential vanishes, there is only one single crossing with two-fold degeneracy as the in-plane exchange coupling increases. (b) At a finite energy, the crossing between the two zero-energy crossings at the finite energy is avoided.   }\label{Mx_cp_vary}
\end{figure}

The absence of the energy crossing of the YSR states (Fig.~\ref{Mx}) at finite energy originates from a finite chemical potential $\mu$ and the in-plane exchange coupling. Since both the out-of-plane and the in-plane exchange couplings commute with $P$, the BdG Hamiltonian at $\mu=0$ can be block-diagonalized. Therefore, the energies of the YSR states in the two blocks are independent. This is the reason for the out-of-plane exchange coupling to produce the crossing at finite energy, which is formed by two YSR states in two different blocks as shown in Fig.~3. However, in the case of an in-plane exchange coupling, the YSR states in the two blocks are degenerate as shown in Fig.~\ref{Mx_cp_vary}(a). Because of the degenerate overlap of the YSR states (one doubly degenerate zero-energy crossing), the crossing at finite energy is irrelevant. Once $\mu$ is moved away from zero, the degeneracy is lifted and the two YSR states form the crossing at finite energy between the two zero-energy crossings. For the out-of-plane exchange coupling, the finite-energy crossing is intact, while the in-plane exchange destroy the crossing as shown in Fig.~\ref{Mx_cp_vary}(b).


\end{document}